\documentclass[dvips,, 12pt]{article}
\usepackage{latexsym, epsfig, amsmath, graphicx, amssymb}
\usepackage{mathrsfs, amsfonts, graphics}
\begin{document}

\begin{center}
\textbf{\Large Foliation of the Kottler-Schwarzschild-De Sitter Spacetime by
Flat Spacelike Hypersurfaces}\\[1.2cm]
Azad A. Siddiqui\\[3ex]

Department of Basic Science and Humanities, EME College, National University
of Science and Technology, Peshawar Road, Rawalpindi, Pakistan,\\[0pt]
E-mail: azad@ceme.nust.edu.pk \\[0pt]
\bigskip

\bigskip

\bigskip

\textbf{Abstract}
\end{center}

\begin{quotation}
There exist Kruskal like coordinates for the Reissner-Nordstrom (RN) black
hole spacetime which are regular at coordinate singularities. Non existence
of such coordinates for the extreme RN black hole spacetime has already been
shown. Also the Carter coordinates available for the extreme case are not
manifestly regular at the coordinate singularity, therefore, a numerical
procedure was developed to obtain free fall geodesics and flat foliation for
the extreme RN black hole spacetime. The Kottler-Schwarzschild-de Sitter
(KSSdS) spacetime geometry is similar to the RN geometry in the sense that,
like the RN case, there exist non-singular coordinates when there are two
distinct coordinate singularities. There are no manifestly regular
coordinates for the extreme KSSdS case. In this paper foliation of all the
cases of the KSSdS spacetime by flat spacelike hypersurfaces is obtained by
introducing a non-singular time coordinate.
\end{quotation}

\section{Introduction}

In General Relativity one often needs slicing of a spacetime by a
sequences of hypersurfaces which is a foliation. A lot of work has
been done on foliation by hypersurfaces of zero mean extrinsic
curvature called \emph{ maximal slicing} \cite{1}-\cite{3} and by
hypersurfaces of constant mean extrinsic curvature or
\emph{CMC-slicing} \cite{4}-\cite{7}. There has been a significant
work to obtain foliation by hypersurfaces of zero intrinsic
curvature called \emph{flat foliation} \cite{8}-\cite{11} as well.
It is known that spherically symmetric static spacetimes admit flat
foliations \cite{10,11} and their uniqueness is also known
\cite{12}. Qadir et.al. have obtained foliations of the
Schwarzschild and RN spacetimes by flat spacelike hypersurfaces
\cite{13} . As the analogue of the Kruskal coordinates does not
exist for the extreme RN spacetime and Carter's coordinates
available for this geometry are not manifestly regular at the
coordinate singularity, a numerical procedure is developed to use
Carter's coordinates to construct free-fall geodesics and a complete
flat foliation of the extreme RN spacetime \cite{14}.

Here we present foliation of the KSSdS by flat spacelike
hypersurfaces (the KSSdS cosmologies and their CMC-slicing has been
discussed and presented in detail in \cite{7}). Instead of following
the procedures similar to those for the RN and extreme RN spacetimes
\cite{13,14}, we have introduced a non-singular time coordinate to
get rid of the coordinate singularities. This removes the coordinate
singularities from the equations giving flat foliating hypersurfaces
and enables us to obtain foliations of all the cases of the KSSdS
spacetime in a much simpler way. In the following section we review
the earlier work to obtain the differential equation satisfied by
flat spherically symmetric hypersurfaces. In Section 3, foliation of
the KSSdS spacetime by flat spacelike hypersurfaces is presented and
in the last section conclusion on our work and some comments on
foliation of the Schwarzschild-anti-de Sitter spacetime by flat
spacelike hypersurfaces are given.

\section{Flat Hypersurfaces Admitted by Spherically Symmetric Static
Spacetimes}

Consider the following spherically symmetric static spacetime metric
\begin{equation}
ds^{2}=-e^{\upsilon \left( r\right) }dt^{2}+e^{\lambda \left( r\right)
}dr^{2}+r^{2}d\Omega ^{2},  \label{1a}
\end{equation}
where
\begin{equation}
d\Omega ^{2}=d\theta ^{2}+\sin ^{2}\theta d\phi ^{2}.  \label{2a}
\end{equation}
Using spherical symmetry to take $\theta $ and $\phi $ constant, an
arbitrary hypersurface in explicit form can be given as
\begin{equation}
t=F(r).  \label{3a}
\end{equation}
The induced 3-metric (of the hypersurfaces) is then
\begin{equation}
ds_{3}^{2}=\left( e^{\lambda \left( r\right) }-e^{\upsilon \left( r\right)
}F^{\prime 2}\right) dr^{2}+r^{2}d\Omega ^{2}.  \label{4a}
\end{equation}
For the induced metric to be flat a \emph{necessary} but \emph{not
sufficient } condition, namely the Ricci scalar, $R=0$, implies
\begin{equation}
\frac{r\left( -\lambda ^{\prime }e^{\lambda }+\nu ^{\prime }e^{\nu
}F^{\prime 2}+2e^{\nu }F^{\prime }F^{\prime \prime }\right) }{\left(
e^{\lambda }-e^{\nu }F^{\prime 2}\right) ^{2}}+\frac{1-e^{\lambda }+e^{\nu
}F^{\prime 2}}{e^{\lambda }-e^{\nu }F^{\prime 2}}=0,  \label{5a}
\end{equation}
where $^{\prime }$ represents the derivative with respect to $r$. Using the
substitution
\begin{equation}
g^{2}\left( r\right) =\frac{1}{e^{\lambda }-e^{\nu }F^{\prime 2}},
\label{6a}
\end{equation}
Eq.$\left( \ref{5a}\right) $ becomes
\begin{equation}
2rgg^{\prime }+g^{2}-1=0,  \label{7a}
\end{equation}
and we have the general solution
\begin{equation}
g^{2}\left( r\right) =1-\frac{k}{r},  \label{8a}
\end{equation}
where $k$ is an arbitrary constant with dimensions of length. The induced
metric now takes the form
\begin{equation}
ds_{3}^{2}=\frac{dr^{2}}{1-\frac{k}{r}}+r^{2}d\Omega ^{2}.  \label{9a}
\end{equation}
The above metric, Eq.$\left( \ref{9a}\right) $, of the hypersurfaces is
flat, i.e. all the components of the Riemann curvature tensor are zero
(which is the necessary and sufficient condition for the hypersurfaces to be
flat), only if $k=0$ or in other words only if $g^{2}\left( r\right) =1$.
Then, from Eqs.$\left( \ref{3a}\right) $ and $\left( \ref{6a}\right) ,$ the
flat spherically symmetric hypersurfaces are uniquely given as \cite{12}
\begin{equation}
t=F\left( r\right) =\int e^{\frac{\lambda -\nu }{2}}\sqrt{1-e^{-\lambda }}dr.
\label{10a}
\end{equation}
The mean extrinsic curvature, $K$, of these hypersurfaces is
\begin{equation}
K=e^{\left( \frac{\nu +\lambda }{2}\right) }\left( \frac{\nu ^{\prime
}e^{\nu }}{2\sqrt{1-e^{\nu }}}-\frac{2\sqrt{1-e^{\nu }}}{r}\right) ,
\label{11a}
\end{equation}
and the Hamiltonian constraint gives
\begin{equation}
R+K^{2}-K_{ab}K^{ab}=\frac{2(K^{2}-e^{\nu })}{r^{2}}-\frac{2\nu ^{\prime
}e^{\nu }}{r},  \label{12a}
\end{equation}
$($here for flat hypersurfaces $R=0)$.

\section{Foliation of the KSSdS Spacetime by Flat Spacelike Hypersurfaces}

The KSSdS metric in gravitational units ($c=G=1$) is given by \cite{7}
\begin{equation}
ds^{2}=-V(r)dt^{2}+V^{-1}(r)dr^{2}+r^{2}(d\theta ^{2}+sin^{2}\theta d\phi
^{2}),  \label{1}
\end{equation}
where
\begin{equation}
V(r)=1-\frac{2m}{r}-\frac{\Lambda r^{2}}{3},  \label{2}
\end{equation}
and the cosmological constant $\Lambda $ and $m$ are positive. In the limit
where $\Lambda $ goes to zero, the spacetime metric tends to the
Schwarzschild metric and in the limit where $m$ goes to zero, the metric
becomes de Sitter.

There are three possible cases depending on the value of $C$, where $C = 9
m^{2}\Lambda$.

\textbf{Case I:} If $C<1$, we call it usual black hole. In this case we have
two horizons, namely the \textit{black hole horizon}, $r_{b}$, and the
\textit{cosmological horizon}, $r_{c}$, which satisfy
\begin{equation}
2m<r_{b}<3m<\frac{1}{\sqrt{{\Lambda }}}<r_{c}<\frac{3}{\sqrt{{\Lambda }}}.
\label{3}
\end{equation}
The function $V(r)$ is zero at these horizons and is positive in the
interval $(r_{b},r_{c})$. The spacetime can be covered by two
coordinate patches: one valid in the region $0<r<r_{c}$ and the
other in the region $ r_{b}<r<\infty $.

\textbf{Case II:} If $C=1$, the \textit{black hole horizon} and the
\textit{ \ cosmological horizon} coincide at $3m$ and we have an
extreme black hole with maximal mass $m=\frac{1}{3\sqrt{\Lambda }}$
and maximal size $ r_{b}=r_{c}=3m$. In this case no non-singular
coordinates are available that remove both singularities
simultaneously.

\textbf{Case III:} If $C > 1$, there are no horizons and we have a naked
singularity case.

Now substituting $e^{\upsilon \left( r\right) }=e^{-\lambda \left(
r\right) }=V(r)$ in Eq.$\left( \ref{10a}\right) $, where $V(r)$ is
given by Eq.$ \left( \ref{2}\right) $, we obtain the equation
satisfied by flat spacelike hypersurfaces for the KSSdS spacetime as

\begin{figure}[tbh]
\begin{center}
\includegraphics[width=.5\textwidth]{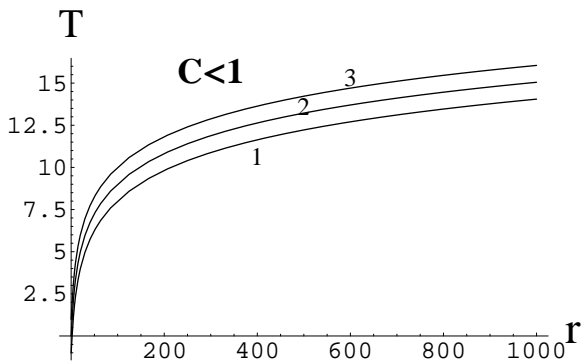}\\[0pt]
\end{center}
\caption{\emph{Flat foliating hypersurfaces in (T, r) coordinates
for $C<1$ (the usual black hole case). The hypersurfaces labeled as
1, 2 and 3 correspond to the values of the foliating parameter,
$T_{0}=-$1, 0 and 1 respectively. $ \Lambda =m=3/7$ for all
hypersurfaces. }}
\end{figure}
\begin{equation}
t=\int \frac{\sqrt{\frac{2m}{r}+\frac{\Lambda
{r^{2}}}{3}}}{\frac{2m}{r}+ \frac{\Lambda {r^{2}}}{3}-1}dr+t_{0},
\label{4}
\end{equation}
where $t_{0}$ is the constant of integration and its different
values correspond to different flat hypersurfaces in ($t,$ $r$)
coordinates. Solving Eq.$\left( \ref{4}\right) $ gives the required
flat foliating spacelike hypersurfaces. It is not difficult to
obtain numerical solution for Case III, as there are no coordinate
singularities. One can also try to obtain results in Case I (like
the usual RN case \cite{13}) by solving Eq.$ \left( \ref{4}\right) $
numerically  separately in two coordinate patches and matching the
solution at a point between $r_{b}$ and $r_{c}$. For the extremal
case one could try to follow the procedure adopted for the extreme
RN case \cite{14}. However, instead of following these procedures to
get rid of the coordinate singularities, we first write Eq.$\left(
\ref{4}\right) $ (by adding and subtracting $1$ in the numerator of
the integral and simplifying) as
\begin{equation}
t=\int \frac{dr}{\sqrt{\frac{2m}{r}+\frac{\Lambda
{r^{2}}}{3}}+1}+\int \frac{ dr}{\frac{2m}{r}+\frac{\Lambda
{r^{2}}}{3}-1}+t_{0},  \label{4b}
\end{equation}
or
\begin{equation}
t=\int \frac{dr}{\sqrt{\frac{2m}{r}+\frac{\Lambda
{r^{2}}}{3}}+1}-\int \frac{ dr}{V(r)}+t_{0}.  \label{4c}
\end{equation}
This motivates to introduce a non-singular time coordinate, $T$, given by
\begin{equation}
dT=dt+\int \frac{dr}{V(r)}.  \label{5}
\end{equation}
Now using Eq.$\left( \ref{5}\right) $ in Eq.$\left( \ref{4c}\right) $ to
obtain the expression for the flat hypersurfaces in ($T,$ $r$) coordinates as

\begin{figure}[tbh]
\begin{center}
\includegraphics[width=.5\textwidth]{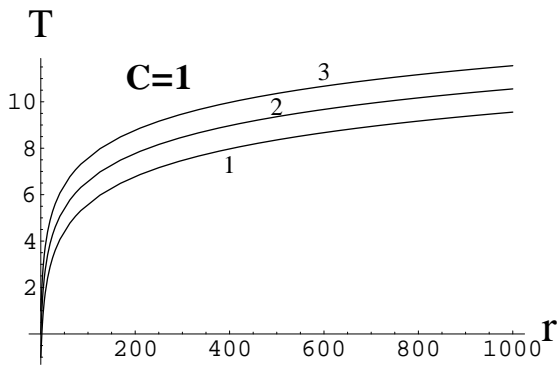}\\[0pt]
\end{center}
\caption{\emph{Flat foliating hypersurfaces in (T, r) coordinates
for $C=1$ (the extreme black hole case). The hypersurfaces labeled
as 1, 2 and 3 correspond to the values of the foliating parameter,
$T_{0}=-$1, 0 and 1 respectively. $\Lambda =1$ and $m=1/3$ for all
hypersurfaces. }}
\end{figure}
\begin{equation}
T=\int \frac{dr}{1+\sqrt{\frac{2m}{r}+\frac{\Lambda {r^{2}}}{3}}}+T_{0},
\label{6}
\end{equation}
where $T_{0}$ is the constant of integration and its different
values give different flat hypersurfaces. The numerical solution of
Eq.$\left( \ref{6} \right) $ for $C$ greater than, equal to and less
than one are obtained and displayed in Figures 1-3 respectively. In
order to see that the results are not artifact of the coordinate
transformation, the numerical solutions of Eq. $\left(
\ref{4}\right) $ are also obtained in Case III (when there are no
coordinate singularities) and displayed in Figure 4. The mean
extrinsic curvature, $K$, of these hypersurfaces is
\begin{equation}
K=\frac{\frac{m}{r^{3}+\frac{2\Lambda
}{3}}}{\sqrt{\frac{2m}{r^{3}+\frac{ \Lambda }{3}}}}.  \label{7}
\end{equation}

\begin{figure}[tbh]
\begin{center}
\includegraphics[width=.5\textwidth]{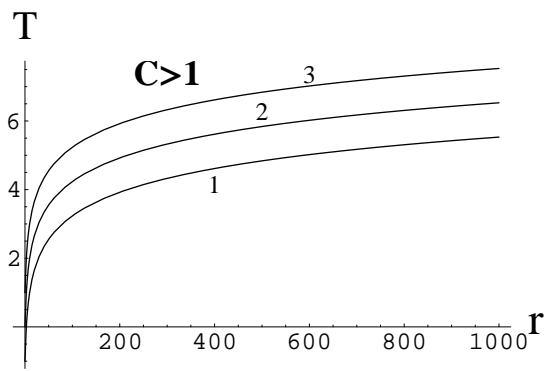}\\[0pt]
\end{center}
\caption{\emph{Flat foliating hypersurfaces in (T, r) coordinates
for $C>1$ (the naked singularity case). The hypersurfaces labeled as
1, 2 and 3 correspond to the values of the foliating parameter,
$T_{0}=-$1, 0 and 1 respectively. $ \Lambda =3$ and $m=1/2$ for all
hypersurfaces. }}
\end{figure}

\section{Conclusion}

Foliation of the RN spacetime was obtained by Qadir et.al. \cite{13}. They
introduced a numerical procedure to deal with the extreme RN case, but the
method was very sensitive near the coordinate singularity and resulted in
the form of kinks in the graphs \cite{14}. Like the RN spacetime, there are
also three cases of the KSSdS spacetime (namely, the usual, extreme and
naked singularity). In this paper foliation of the KSSdS spacetime by flat
spacelike hypersurfaces is obtained by introducing a coordinate
transformation that removes the singularity from the equation of the flat
hypersurfaces in all the cases. In case of the naked singularity we have
obtained flat hypersurfaces both in the original $(t,r)$ coordinates and in
the transformed $(T,r)$ coordinates. The results show that our procedure
works well and the hypersurfaces obtained in all the cases are not artifact
of the transformation. It will be interesting to apply our procedure to the
RN spacetime and compare the results with the earlier results. It is
expected that our procedure will remove the kinks appearing in the graphs
obtained earlier \cite{14}.

\begin{figure}[htb]
\begin{center}
\includegraphics[width=.5\textwidth]{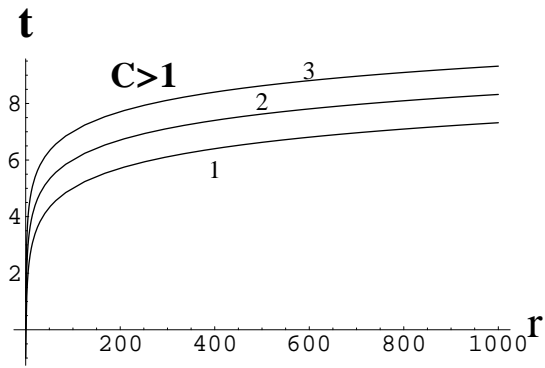}\\[0pt]
\end{center}
\caption{\emph{Flat foliating hypersurfaces in (t, r) coordinates
for $C>1$ (the naked singularity case). The hypersurfaces labeled as
1, 2 and 3 correspond to the values of the foliating parameter,
$t_{0}=-$1, 0 and 1 respectively. $ \Lambda=3$ and $m=1/2$ for all
hypersurfaces. Notice that the behaviour of the corresponding
hypersurfaces in (T, r) coordinates in Figure 3 is essentially the
same. This shows that the results obtained in (T, r) coordinates are
not artifact of the coordinate transformation. }}
\end{figure}

The Schwarzschild-anti-de Sitter spacetime has the same form as given by
Eqs. $\left( \ref{1}\right) $ and $\left( \ref{2}\right) $ for the KSSdS
spacetime, but the cosmological constant, $\Lambda $, now takes negative
values \cite{15}. Replacing $\Lambda $ by $-\Lambda $ and following the same
procedure as discussed in Section 3 for the KSSdS spacetime, we obtain the
following expression for the flat hypersurfaces admitted by the
Schwarzschild-anti-de Sitter spacetime
\begin{equation}
T=\int \frac{dr}{1+\sqrt{\frac{2m}{r}-\frac{\Lambda {r^{2}}}{3}}}+T_{0},\ \
(\Lambda >0).  \label{8}
\end{equation}
Notice that the term inside the square root becomes negative for
$r>\left( \frac{6m}{\Lambda }\right) ^{1/3}$, and restricts the
solution in that region. Therefore, a direct application of our
procedure does not work for the Schwarzschild-anti-de Sitter
spacetime. However, it will be interesting to try to construct some
other non-singular coordinates in this case and also explore the
significance of the barrier at $r=\left( \frac{6m}{\Lambda } \right)
^{1/3}$.

\section{Acknowledgments}

The author is grateful to NUST for the financial support provided to
participate in the 2nd Italian-Pakistani workshop and is also thankful to
Prof. Robert Beig and Khalid Saifullah for useful discussions on the work.
He is also thankful to the unknown referees for their comments and
suggestions that have improved the manuscript significantly.

\end{document}